\documentclass[conference]{IEEEtran}
\IEEEoverridecommandlockouts

\usepackage{graphicx}
\usepackage{amsfonts}
\usepackage{amssymb}

\usepackage{float}
\usepackage{amsmath}
\usepackage{amsfonts}

\usepackage{cite}
\usepackage{multirow}
\usepackage{multicol}
\usepackage[linesnumbered,ruled,vlined, lined, commentsnumbered, longend]{algorithm2e} 
\usepackage{amsmath,amssymb,amsfonts}
\usepackage{algorithmic}
\usepackage{graphicx}
\usepackage{textcomp}
\usepackage{xcolor}

\usepackage{cite}
\usepackage{multirow}
\usepackage{multicol}

\usepackage[linesnumbered,ruled,vlined, lined, commentsnumbered, longend]{algorithm2e} 

\usepackage{etoolbox}

\usepackage{amsmath,amssymb,amsfonts}

\usepackage{xcolor}
\usepackage[colorlinks = true,
            linkcolor = blue,
            urlcolor  = blue,
            citecolor = blue,
            anchorcolor = blue]{hyperref}
\usepackage{orcidlink}

\usepackage{scalerel}
\usepackage{tikz}
\usetikzlibrary{svg.path}
\usepackage{scalerel}[2016/12/29]

\definecolor{orcidlogocol}{HTML}{A6CE39}

\tikzset{
  orcidlogo/.pic={
    \fill[orcidlogocol] svg{M256,128c0,70.7-57.3,128-128,128C57.3,256,0,198.7,0,128C0,57.3,57.3,0,128,0C198.7,0,256,57.3,256,128z};
    \fill[white] svg{M86.3,186.2H70.9V79.1h15.4v48.4V186.2z}
                 svg{M108.9,79.1h41.6c39.6,0,57,28.3,57,53.6c0,27.5-21.5,53.6-56.8,53.6h-41.8V79.1z M124.3,172.4h24.5c34.9,0,42.9-26.5,42.9-39.7c0-21.5-13.7-39.7-43.7-39.7h-23.7V172.4z}
                 svg{M88.7,56.8c0,5.5-4.5,10.1-10.1,10.1c-5.6,0-10.1-4.6-10.1-10.1c0-5.6,4.5-10.1,10.1-10.1C84.2,46.7,88.7,51.3,88.7,56.8z};
  }
}

\newcommand\orcid[1]{\href{https://orcid.org/#1}{\mbox{\scalerel*{
\begin{tikzpicture}[yscale=-1,transform shape]
\pic{orcidlogo};
\end{tikzpicture}
}{|}}}}

\def\BibTeX{{\rm B\kern-.05em{\sc i\kern-.025em b}\kern-.08em
    T\kern-.1667em\lower.7ex\hbox{E}\kern-.125emX}}
\begin{document}

\title{Generative Adversarial Network based Voice Conversion: Techniques, Challenges, and Recent Advancements
}

\author{\IEEEauthorblockN{Sandipan Dhar$^1$, Nanda Dulal Jana$^1$, Swagatam Das$^2$}
    \IEEEauthorblockA{$^1$\textit{Department of Computer Science and Engineering, National Institute of Technology Durgapur, India.}}
    \IEEEauthorblockA{$^2$\textit{Electronics and Communication Sciences Unit, Indian Statistical Institute, Kolkata, India.}}
    \IEEEauthorblockA{Email: \url{sandipandhartsk03@gmail.com}, \url{ndjana.cse@nitdgp.ac.in}, \url{swagatam.das@isical.ac.in}}}

\maketitle

\begin{abstract}
Voice conversion (VC) stands as a crucial research area in speech synthesis, enabling the transformation of a speaker's vocal characteristics to resemble another while preserving the linguistic content. This technology has broad applications, including automated movie dubbing, speech-to-singing conversion, and assistive devices for pathological speech rehabilitation. With the increasing demand for high-quality and natural-sounding synthetic voices, researchers have developed a wide range of VC techniques. Among these, generative adversarial network (GAN)-based approaches have drawn considerable attention for their powerful feature-mapping capabilities and potential to produce highly realistic speech. Despite notable advancements, challenges such as ensuring training stability, maintaining linguistic consistency, and achieving perceptual naturalness continue to hinder progress in GAN-based VC systems. This systematic review presents a comprehensive analysis of the voice conversion landscape, highlighting key techniques, key challenges, and the transformative impact of GANs in the field. The survey categorizes existing methods, examines technical obstacles, and critically evaluates recent developments in GAN-based VC. By consolidating and synthesizing research findings scattered across the literature, this review provides a structured understanding of the strengths and limitations of different approaches. The significance of this survey lies in its ability to guide future research by identifying existing gaps, proposing potential directions, and offering insights for building more robust and efficient VC systems. Overall, this work serves as an essential resource for researchers, developers, and practitioners aiming to advance the state-of-the-art (SOTA) in voice conversion technology.
\end{abstract}

\begin{IEEEkeywords}
Voice Conversion, Collective Learning Mechanism, Optimal Transport Loss, Conformer, Generative Adversarial Network.
\end{IEEEkeywords}

\section{Introduction}
\label{chap:intro}
\begin{figure*}[htbp]
    \centering
    \includegraphics[height=4cm, width=12cm]{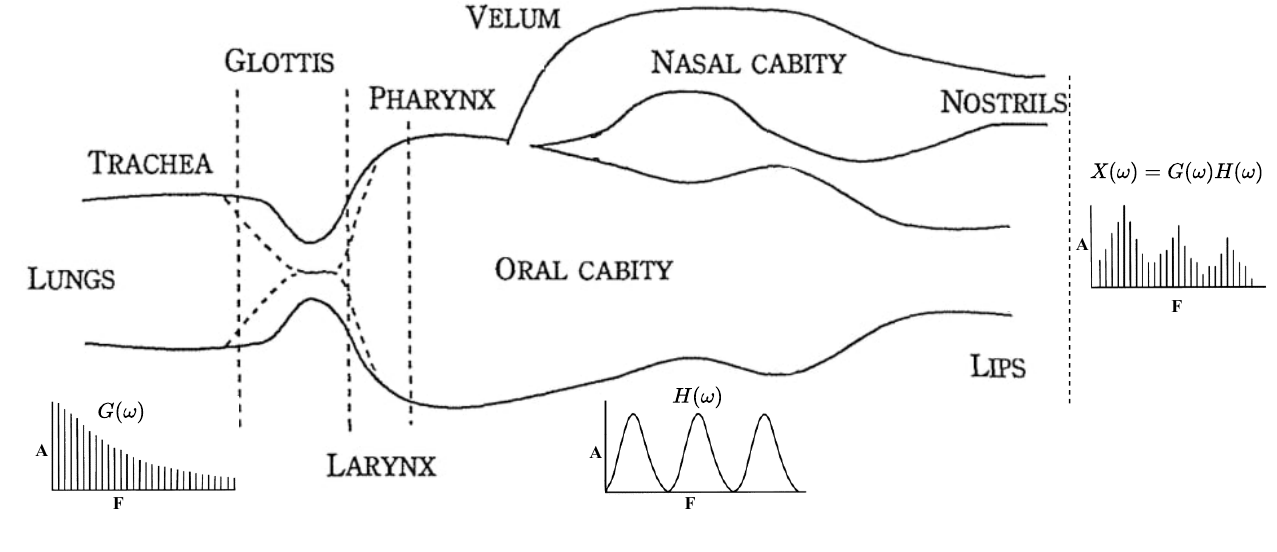}
    \caption{{Schematic overview of the human speech production mechanism}}
    \label{fig:thesis-1}
\end{figure*}

Speech refers to the process of producing meaningful vocalized sounds that carry linguistic information for verbal communication. In natural verbal communication between a talker and a listener, the talker plays the role of speech generator while the listener takes on the task of speech recognition. Digital speech processing (DSP) is a sub-field of signal processing that analyses the physical aspects of speech as digital signals \cite{Berrak_Sisman}. In DSP, the process of speech generation can be described as follows:
\par
\begin{equation}
    x(t)=\int^{\infty}_{-\infty} g(\tau) h(t-\tau) d\tau,
    \label{Eq.1}
\end{equation}
where $g(\tau)$ is the excitation signal ($\tau$ is variable), $h(t-\tau)$ is the transformation function (vocal tract transformation function), and $x(t)$ is the output speech signal which is a time-varying signal \cite{Thomas_Quatieri}. In frequency domain Eq. (\ref{Eq.1}) can be represented as:
\begin{equation}
    X(\omega)= G(\omega) H(\omega).
    \label{Eq.2}
\end{equation}
Here in Eq. (\ref{Eq.2}), $G(\omega)$ ($\omega$ implies angular frequency) is the frequency response of the excitation signal, $H(\omega)$ is the frequency response of the transformation function  (i.e., the frequency response of the vocal tract transformation function) and $X(\omega)$ is the frequency response of the output speech signal \cite{Holmes_Wendy}. In Fig. \ref{fig:thesis-1}, the schematic overview of the human speech production mechanism (source filter model) is shown, which is based on Eq. (\ref{Eq.2}). 
\par
\par
In contrast to the natural speech generation process, speech synthesis is the process of artificially generating human speech either by statistical models or by artificial intelligence (AI) based algorithms \cite{Jurafsky}. Speech synthesis deals with understanding speaker-dependent characteristics, such as vocal features, speaking style, emotional conditions, etc., for generating new speech samples given any specific speech content \cite{Speech_Synthesis} (speech contents are extracted using automatic speech recognition models, i.e. ASR models). The primary goal in a speech synthesis process is to analyze $X(\omega)$ associated with a particular speech content or phonetic content (e.g., some phonetic content c) for understanding the characteristics of $H(\omega)$. This can be interpreted in terms of a mapping function $M$, which maps a phonetic content $c$ (that belongs to a complete time interval $t$) to its corresponding audible speech signal $x(t)$ (i.e., $c\implies M \implies x(t)$). This understanding helps in generating new audible speech waveforms $\hat{x}(t)$ when presented with new speech content $\hat{c}$ (i.e., $\hat{c}\implies M \implies \hat{x}(t)$). Nevertheless, the main challenge in speech synthesis lies in achieving human-like, natural-sounding speech \cite{Neural_Speech_Synthesis}.
\par
\begin{figure}[htbp]
    \centering
    \includegraphics[height=1.8cm, width=6.8cm]{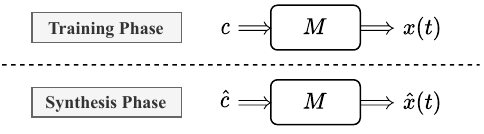}
    \caption{Overview of the training and synthesis phase of the speech synthesis process}
    \label{fig:thesis-0001}
\end{figure}
\begin{figure*}[htbp]
    \centering
    \includegraphics[height=8.5cm, width=15.5cm]{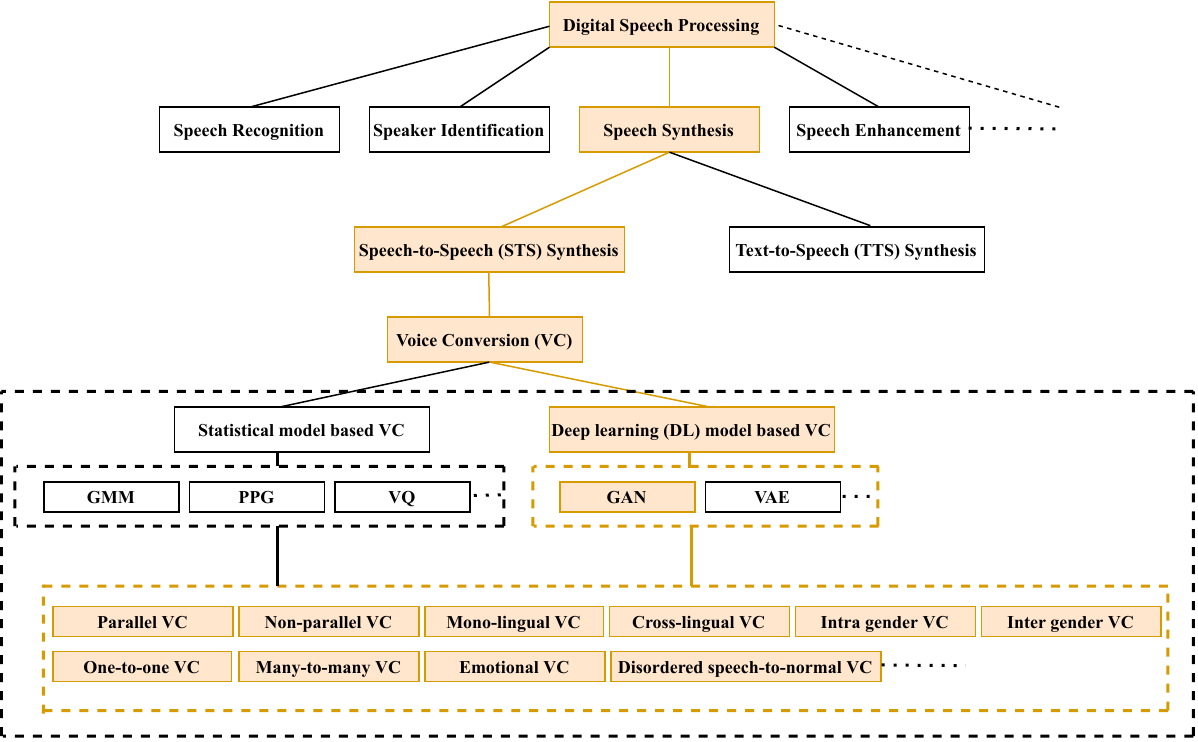}
    \caption{Categorization of the speech synthesis processes}
    \label{fig:thesis-001}
\end{figure*}

The speech synthesis process consists of two phases: training and synthesis. During the training phase, a mapping module, which can be either statistical or AI-based, attempts to associate extracted speech features from input with their corresponding audible speech. The trained mapping module or mapping function generates new audible speech conditioned on new input in the synthesis phase. Based on the type of mapping technique and the type of input and output data, the speech synthesis systems are divided into two main categories: text-to-speech (TTS) synthesis \cite{TTS} and speech-to-speech (STS) synthesis systems \cite{Speech_synthesis_review}. Fig. \ref{fig:thesis-0001} 
shows the overview of the training and synthesis phase of a typical TTS synthesis process. TTS and STS synthesis systems are further divided into multiple sub-categories. Despite the fact that the TTS is a well-established domain of research, its major limitation is its need for both speaker-independent (i.e., speech content in terms of textual transcript) and its corresponding speaker-dependent (i.e., audible speech data) information for the training purpose. Thus, it works efficiently for seen speakers rather than unseen speakers. Meanwhile, in a STS synthesis system, the emphasis is more on speaker-related information rather than its content information (as the content information mostly remains unaltered in a STS synthesis process). Thus, in recent years, voice conversion (VC) \cite{Berrak_Sisman} emerged as an exponentially growing field of research that comes under the paradigm of STS synthesis, as depicted in Fig. \ref{fig:thesis-001}.
\par
\begin{figure}[htbp]
    \centering
    \includegraphics[height=3cm, width=8.0cm]{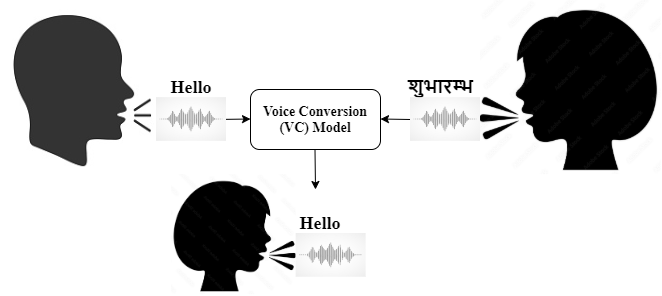}
    \caption{Schematic overview of the VC process}
    \label{fig:thesis-2}
\end{figure}

VC is the process of converting the vocal texture of a source speaker to that of a target speaker, keeping the linguistic content of the source speaker unaltered \cite{Berrak_Sisman}. It implies that in a VC process, speaker-dependent information is transferred from the target to the source domain without altering the speaker-independent information of the source part (as the speaker-independent information remains the same throughout the process) \cite{speech_recognition_survey}. The schematic overview of the VC process is shown in Fig. \ref{fig:thesis-2}. Recently, VC has emerged as a continuously evolving domain within the speech synthesis research, presenting promising applications across various real-time contexts \cite{Berrak_Sisman}. Numerous efforts have been undertaken to effectively execute the vocal style transformation task, spanning from statistical approaches to generative deep learning models.
\par
The VC process can be categorised into different types based on the linguistic contents of the speech datasets and the type of VC mechanism employed. The most prominent types of VC processes are parallel/non-parallel VC, mono-lingual/cross-lingual VC, intra-gender/inter-gender VC, one-to-one/many-to-many VC, emotional VC, disordered speech-to-normal VC, singing VC (SVC), etc. \cite{Berrak_Sisman}\cite{VC-Overview}.
\par
\begin{figure}[htbp]
    \centering
    \includegraphics[height=2.05cm, width=8.5cm]{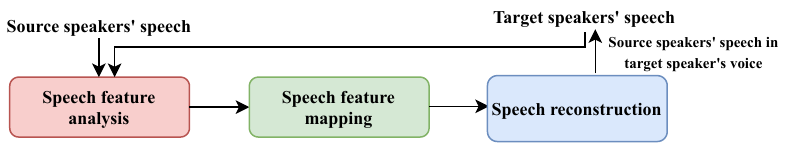}
    \caption{Three-stage pipeline of the VC process}
    \label{fig:thesis-3}
\end{figure}
\begin{figure*}[htbp]
    \centering
    \includegraphics[height=6cm, width=14.5cm]{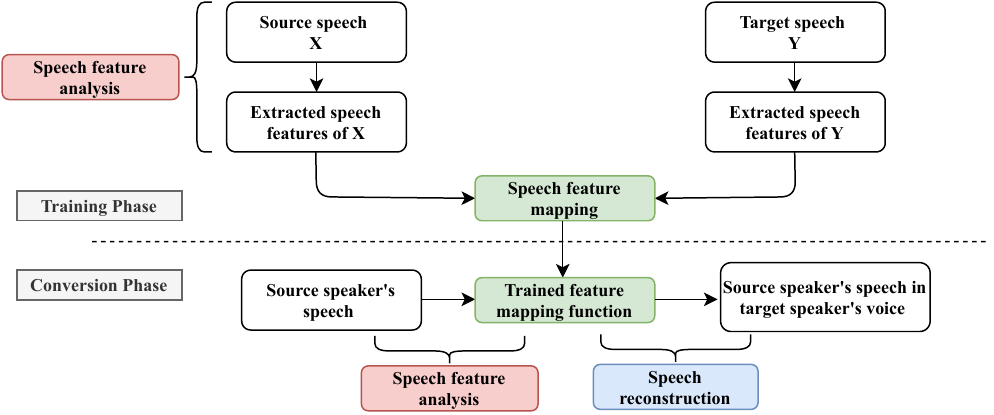}
    \caption{Implementation of the three-stage pipeline in the training and conversion phase of the VC process}
    \label{fig:thesis-4}
\end{figure*}

A VC process follows a three-stage pipeline: speech feature analysis, mapping, and reconstruction \cite{An_overview_of_VC}. Fig. \ref{fig:thesis-3} illustrates this three-stage pipeline of a VC process.  
During the speech feature analysis stage, the speech samples from both the source and target speakers are analyzed to extract specific features representing the segmental and supra-segmental information. In the subsequent speech feature mapping stage, a mapping function is employed to convert the source speakers' vocal features similar to that of a target speaker. This mapping function is crucial in efficiently transferring the vocal characteristics from the source to the target speaker without altering the speech content. Finally, the converted speech features are utilized to reconstruct the audible speech in the speech reconstruction stage. In the speech reconstruction stage, statistical or neural vocoders are used to generate reconstructed audible speech sample containing the source speakers' utterance in the target speaker's voice. Like any other traditional speech synthesis process, the VC module is also implemented following two phases: the training phase and the conversion phase. In Fig. \ref{fig:thesis-4}, the working mechanism of the three-stage pipeline in the training and conversion phase is shown by considering source speaker $X$ and target speaker $Y$. 
\par
\subsection{Speech Feature Analysis}
This three-stage pipeline of the VC process can also be expressed mathematically for better interpretation. As depicted in Eq. (\ref{eq:Equation11}), ${F(.)}$ represents the speech analysis function (a composite function) consisting of ${k}$ speech feature extraction functions that pull out the speech features ${\mathbf{x}}$ from the source speaker's (denoted as ${X}$) audible speech sample ${\mathbf{s}}$,
\begin{equation}
\begin{split}
{{F(.)=f_k(...(f_3(f_2(f_1(.))))),}}
\label{eq:Equation11}
\end{split}
\end{equation}
\begin{equation}
\begin{split}
{{\mathbf{x}=F(\mathbf{{s}})}.}
\label{eq:Equation012}
\end{split}
\end{equation}In the speech feature analysis stage \cite{Speech_Analysis}, the spectral and prosodic features are extracted. The spectral features mostly considered in VC are spectral envelope, mel-cepstral coefficients (MCC), modulation spectra (MS), mel-frequency cepstral coefficients (MFCC), aperiodicity (AP) of speech, $F_{0}$-fundamental frequency, to name a few \cite{VC_Analysis} \cite{Speech_Features}. Apart from these, the mel-spectrogram \cite{Speech_Features_1} is also a well-known speech feature representation widely used in the domain of VC. The speech features extracted in the speech analysis stage play a significant role in the overall speech synthesis process. However, the speech feature analysis stage highly depends on the vocoders \cite{Vocoder} used in the speech reconstruction phase, as audible speech must be reconstructed from the extracted speech features (which is the inverse of the speech analysis stage).
\subsection{Speech Feature Mapping}
 In the speech feature mapping stage, different statistical and deep learning (DL) algorithms are used for source-to-target speech feature mapping \cite{Berrak_Sisman}. In Eq. (\ref{eq:Equation22}), ${G(.)}$ is the speech feature mapping function that maps the speech features from the source domain $X$ to the target domain $Y$ and ${{\mathbf{\hat{y}}}}$ represents the converted speech features, expressed as follows:
\begin{equation}
{{\mathbf{\hat{y}}=G(\mathbf{x})}.}
\label{eq:Equation22}
\end{equation}During the initial phase of the development of VC models, extensive research was conducted considering statistical algorithms like Gaussian mixture models (GMM) \cite{GMM-VC}, phonetic-posteriorgram (PPG) \cite{PPG-VC}, and vector quantization (VQ) models \cite{VQ-VC}, etc. \cite{Berrak_Sisman, VC-Overview}, for speech feature mapping.
\par
In GMM-based VC \cite{GMM-VC, GMM-VC-1, GMM-VC-2}, a joint vector $\mathbf{z}$ is formed after aligning the source and the target speech feature vectors $\mathbf{x}$, $\mathbf{y}$, such that $\mathbf{z=[x^{T},y^{T}]}$. During training, the aligned data $\mathbf{z}$ is used to estimate the parameters ({$\mathbf{\alpha}$, i.e., the weight associated with a mixture component, mean vector $\mathbf{\mu}$, covariance matrix $\mathbf{\Sigma}$}) of the GMM for $Q$ Gaussian components iteratively, by using the expectation maximization (EM) algorithm. After training, the mapping function $G(.)$ is learned using the minimum mean squared error (MMSE). The MMSE estimator estimates mapping function $G(.)$, such that $\hat{\mathbf{y}}=G(\mathbf{x})$. MMSE estimator can be represented in terms of conditional expectation $\mathbb{E}[y \mid x]$. Given the $i^{th}$ Gaussian components (i.e., $\mathbf{\mu_{i}}$, $\mathbf{\Sigma_{i}}$), the mapping function $G(.)$ can be computed as, 
\begin{equation}
\hat{{y_i}}=G({x_i})
=\mathbb{E}[y \mid x]
=\sum^{Q}_{i=1} h_{i}(x)[\mu^{y}_{i}+\sum^{yx}_{i}(\sum^{xx}_{i})^{-1}(x-\mu^{x}_{i})],
\label{05}
\end{equation}
where $h_{i}(x)$ is the probability that the $i^{th}$ Gaussian component generated the
vector $x$. However, GMM faces significant problems, such as over-fitting and over-smoothing, which results in losing the fine details of the spectrum \cite{over-smoothing-GMM-based}.
\par
Meanwhile, PPG is a specific representation of the spoken phrase where the horizontal axis represents time, and the vertical one contains indices of phonetic classes \cite{PPG-VC, PPG-1, PPG-2}. PPGs $(P_1, P_2,..., P_N)$ are obtained using speaker-independent automatic speech recognition (SI-ASR) models in PPG-based VC. Sequential models are utilized to obtain the speech feature sequence $(\hat{Y}_1, \hat{Y}_2,..., \hat{Y}_N)$ from the PPGs to compare with the actual target sequence $({Y}_1, {Y}_2,..., {Y}_N)$. The model parameters are trained by minimizing errors between the obtained speech feature sequence and the target sequence. However, one major limitation of the PPG-based methods is that the effectiveness of the models heavily relies on the adequate representation of the training data and thus often struggles to adapt to extreme variations in voice characteristics. Leveraging the progress made in representing the phonetic information through PPGs, various hybrid models are employed as the mapping module to perform VC. 
\par
On the other hand, in the VQ technique-based VC mapping \cite{VQVC} \cite{VQ-VC}, mapping codebooks are generated where the spoken words of two speakers are vector-quantized, and thus the continuous data ($V$ as a sequence of continuous data, i.e., $V=v_1, v_2, ...v_T$) is transformed into discrete data ($VQ(V) = q_1, q_2,...,q_T$, i.e., discrete representation of the continuous data). Then, the dynamic time warping (DTW) algorithm \cite{DTW} is used to find the correspondence between the vectors of the same words uttered by both speakers. The vector correspondences between the speakers are accumulated as histograms, representing the linear combination of the target speaker's vector. Mapping codebooks are also generated for pitch frequencies and power values (these are scalar quantized instead of vector quantized). Finally, in the conversion phase, a synthesis filter is used to reconstruct the speech in the target speakers' vocal features. Apart from the discussed VC methods, there are other existing techniques, such as non-negative matrix factorization (NMF) \cite{NMF}, dynamic kernel partial least squares (DKPLS) \cite{DKPLS}, hidden markov model (HMM) \cite{HMM-based-VC} and more \cite{Berrak_Sisman}. 
\par
However, with the flourishing development in the field of generative AI, deep generative models such as variational autoencoders (VAEs) \cite{VAE}, generative adversarial networks (GANs) \cite{GAN}, etc. \cite{VC_DeepGenerativemodel, Berrak_Sisman}, emerged as a better alternative to their statistical counterparts due to their robust nature, realistic data generation capability and efficient domain adaptability, as depicted in Fig. \ref{fig:thesis-001}. Moreover, the generated speech samples of statistical VC models vastly need to improve on the over-smoothing problem, which has given rise to the utilisation of DL-based models for resolving these limitations of conventional models. Based on the preceding literature, it is evident that due to the impressive capability to generate realistic data in image style transfer research, GAN models have emerged as a significant choice within the VC research community for performing the speech synthesis task \cite{GAN_1, Shah_Nirmesh6}. The introduction of GAN models marked as a novel approach in VC research for improving the quality and naturalness of the generated speech samples \cite{A-Survey-VC}. The field of deep generative model-based VC research is continually evolving due to the progress made to the frameworks of GAN models. Furthermore, GANs have several advantages over the other existing variants of generative models, such as VAEs \cite{GAN-vs-VAE, GAN-vs-VAE_1, GAN-vs-VAE_2}. One notable benefit of GAN is the adversarial learning strategy, which enhances its ability to capture the target distribution more effectively. The GAN model learns the target distribution based on the indirect supervision of the discriminator, which provides continuous feedback to the generator model such that it can improve its data generation ability. This learning mechanism makes GAN more powerful in the context that it can get an indirect estimation of the target distribution in its initial stage of training only, which is missing in preceding generative models such as VAEs or auto-regressive models. In this context, many variants of GAN models have been developed to perform the VC task efficiently \cite{Variantsof_GAN}. Among these, variants of CycleGAN-VC \cite{CycleGAN-VC} have shown significant performance for efficient transfer of speech characteristics as one of the initial implementations of GAN-based VC. The utilisation of GAN models in the field of VC has introduced a new research direction of deep generative model-based STS synthesis. 
\par
\subsection{Speech Reconstruction}
\begin{figure}[h]
    \centering
    \includegraphics[height=3.75cm, width=7.0cm]{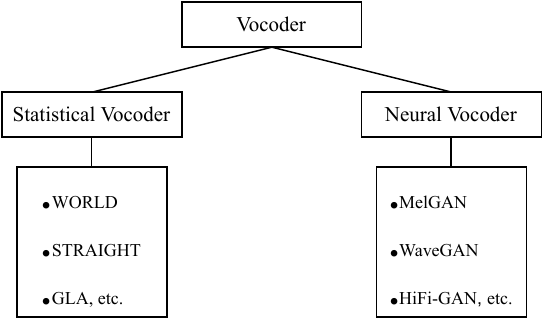}
    \caption{Types of statistical and neural vocoders}
    \label{fig:thesis-vocoder}
\end{figure}

Apart from speech feature analysis and mapping, the speech reconstruction stage also holds substantial importance in determining the quality of the generated speech samples and is mainly dependent on the choice of vocoders \cite{vocoder-121, Neural_Vocoder}, i.e., statistical and neural vocoders, as shown in Fig. \ref{fig:thesis-vocoder}. The speech reconstruction function ${F^{-1}(.)}$ (i.e., vocoder) reconstructs the audible speech in target speaker Y's voice, i.e. ${F^{-1}(.)}$ is used to obtain ${\mathbf{\hat{t}}}$ from speech feature ${\mathbf{\hat{y}}}$ which is the inverse operation of speech analysis function ${F(.)}$ and can be expressible as:
\begin{equation}
\begin{split}
{{\mathbf{\hat{t}}=F^{-1}(\mathbf{\hat{y}})},}
\label{eq:Equation3}
\end{split}
\end{equation}such that it sounds like the target speaker is uttering the speech content of the source speaker. In GAN-based VC, several prominent statistical vocoders are employed, such as WORLD \cite{World_Vocoder}, STRAIGHT \cite{STRAIGHT_Vocoder}, Griffin-Lim algorithm (GLA) \cite{Griffin-Lim_Algorithm}, etc \cite{Shah_Nirmesh5}. Most of these statistical vocoders depend on multiple speech features to reconstruct audible speech.  On the other hand, neural vocoders (i.e. neural network-based vocoders) such as MelGAN \cite{melgan}, WaveGAN \cite{Wavegan}, HiFi-GAN \cite{HiFi-GAN}, MFCCGAN \cite{MFCCGAN}, etc., predominantly rely on single speech features such as mel-spectrogram, MFCC, etc. \cite{Speech_Features_Information_Fusion}, for the reconstruction of speech in audible format. The selection of vocoders significantly impacts the generated quality of the speech samples. 
Neural vocoders offer several significant advantages over traditional statistical vocoders in the speech synthesis process. Firstly, they produce higher audio quality, capturing the nuances of human speech with greater fidelity, which results in more natural-sounding outputs. Additionally, neural vocoders excel in maintaining temporal coherence, leading to smoother transitions between phonemes and a more fluid listening experience. They are particularly adept at modelling complex waveforms, allowing for richer and more varied speech synthesis. Their adaptability is another key benefit; they can easily adjust to different speakers and styles, making them versatile for personalized voice applications. Moreover, neural vocoders typically generate fewer perceptible artifacts compared to statistical methods, resulting in clearer and more pleasant audio. Furthermore, neural vocoders can incorporate prosody and emotional variations effectively, enhancing expressiveness in synthesized speech. However, statistical vocoders also have many advantages. It can provide consistent, deterministic outputs and typically requires less computational power, enabling real-time processing in resource-constrained environments. Their parameters are easier to interpret, allowing for direct control over synthesis aspects like pitch and duration. Additionally, they can work effectively with smaller datasets. Overall, their long history of development and compatibility with legacy systems make statistical vocoders a valuable option in speech synthesis applications.
\section{Issues \& Challenges in VC}
In recent years, DL models have significantly contributed to the STS synthesis VC research, primarily in developing various human interactive audio assistive AI applications \cite{Berrak_Sisman}. These applications encompass various domains, ranging from voice dubbing for films and television to automatic speech synthesis solutions for individuals with speech disorders, empowering AI-driven virtual assistants capable of verbal communication. Despite the remarkable advancements, speech synthesis research grapples with the formidable challenge of generating natural-sounding speech that captures human speech's intricate nuances and realistic properties \cite{Speech_Synthesis}. Unlike speaker-independent features like textual transcripts, which are merely a sequence of words, audible human speech is associated with emotion, intonation, rhythm, accent, etc., related to individual speakers. Therefore, replicating these aspects convincingly requires sophisticated algorithms to analyze the subtle variations in speech patterns. This persistent challenge underscores the critical need for exploring and developing appropriate solutions within the realm of DL-based VC research. Traditional VC models often struggle to emulate this variability of speech patterns, leading to synthesized speech that sounds robotic or unnatural \cite{Neural_Speech_Synthesis}. Hence, this area represents a crucial frontier in STS synthesis research, demanding concerted efforts and extensive studies to unravel its complexities and devise innovative solutions. Though significant progress has been made, still there remains a noticeable gap in research focusing on achieving realistic synthesized human speech using deep generative models like GANs. Most of the recent advancements in GAN-based VC research predominantly concentrate on building various variants of GAN frameworks to cater to a broad spectrum of applications, spanning from one-to-one \cite{One-to-One} to many-to-many VC \cite{Variantsof_GAN} scenarios. However, the other associated factors, such as loss functions, learning mechanisms, training procedures, the utilization of different speech features, etc., have been less explored, all of which significantly contribute to developing efficient GAN models capable of effectively handling diverse VC processes. Furthermore, the field of speech research faces a substantial challenge due to the scarcity of audible speech data, particularly concerning low-resource languages and pathological speech samples. Utilizing proficient GAN models capable of generating highly realistic augmented speech samples offers a potential solution to this data scarcity problem.
\par
Several prominent research challenges, as previously discussed, continue to affect GAN-based VC. These include generating high-quality realistic speech samples, exploring suitable DL frameworks and speech feature-specific loss functions, improvising different learning mechanisms, selecting appropriate vocoders for generating synthesized speech samples with high speaker similarity, etc. Each of these challenges plays a crucial role in shaping the effectiveness of GAN-based VC systems. These key research challenges are elaborately discussed as follows:
\begin{itemize}
    \item {Proficient generation of high-quality natural-sounding speech data for diverse applications related to VC. Specifically tackling the problem of limited data availability across various low-resource languages, speaking styles, and pathological speech conditions.}
    \item {Finding the impact of diverse DL architectures and their associated components in the GAN models, deployed for capturing the subtle nuances of natural human speech across a variety of VC processes.}
    \item {Exploring the effect of various loss functions on the training process of GAN-based VC models to improve the speech quality of the generated speech samples, ensuring the preservation of the linguistic content.}
    \item {Assessing the contribution of different speech features and vocoders to effectively capture the target feature distribution by end-to-end feature comparison technique, to facilitate the efficient mapping of speech features from the source to the target domain by obtaining high vocal feature similarity in terms of both objective and subjective measurements.}
    \item {Analysing the impact of different learning mechanisms in the GAN training process and the improvisation of the learning strategies to make speech synthesis approaches more robust and adaptive.} 
\end{itemize}

\section{Background and Literature}
\subsection{GAN-based VC}
The GAN model is one of the most promising models among the class of generative models \cite{GAN}. These deep generative models are widely used for synthetic data generation in various applications, initially implemented in the field of computer vision and later used for speech synthesis.  A typical GAN model consists of a generator and a discriminator, denoted by $G$ and $D$. The generator generates synthetic data from prior input information, whereas the discriminator distinguishes the generated synthetic data from the original training data \cite{GAN-Survey-CV}. Consider the set of model parameters of a typical generator and a discriminator are $\theta_{g}$ and $\theta_{d}$, respectively. Given an input data $\textbf{x}$, the generator generates synthetic or fake data $\textbf{\^y}$ having a probability distribution ${p_{\textbf{\^y}}}$. On the other hand, the discriminator is trained with the original or real data $\textbf{y}$ possessing a probability distribution ${p_\textbf{y}}$. The generator tries to map the input data from domain ${X}$ to ${Y}$. For the generated data $\textbf{\^y}$, the discriminator outputs a probability value or score, which indicates the probability of belongingness of $\textbf{\^y}$ to the real or fake class. The generator's objective in GAN is to deceive the discriminator by generating more real-like synthetic data. On the other hand, the discriminator's objective is to enhance its ability to classify fake samples as fake and real samples as real. Thus, both the discriminator and the generator learn in an adversarial manner by maximizing and minimizing the objective function or loss function defined as,
\vspace{-0.2cm} 
\begin{equation}
\begin{split}
\mathcal{L}_{CEL}(\mathbf{y}, \mathbf{\hat{y}} ) = \underset{G}{\mathrm{min}}\ \underset{D}{\mathrm{max}} \Bigg[ 
&\ \mathbb{E}_{\mathbf{y} \sim p_{\mathbf{y}}} \big( \log D_{\theta_d}(\mathbf{y}) \big) \\
+ &\ \mathbb{E}_{\mathbf{x} \sim p_{\mathbf{x}}} \big( \log (1 - D_{\theta_d}(G_{\theta_g}(\mathbf{x}))) \big) 
\Bigg],
\end{split}
\label{0}
\end{equation}
\vspace{-0.1cm} 
where CEL denotes \cite{CEL} the Cross-Entropy Loss and $\mathbb{E}$ is expectation. The model parameters $\theta_{d}$ and $\theta_{g}$ are updated through maximization and minimization of the loss function. This minimization and maximization of the common loss function originate from a two-player zero-sum game \cite{zero-sum-gam} between the generator and the discriminator. As the GAN training progresses, the probability distribution ${p_\textbf{\^y}}$ is expected to become similar to ${p_\textbf{y}}$, thereby resulting in an equilibrium state known as the Nash equilibrium \cite{GAN-Survey-2}. In this state, the discriminator outputs a score of 0.5 (probability value) for the synthesized data, which implies an equal probability of belongingness to a real or fake class. This indicates that the generated samples are indistinguishable from their original counterpart.  
\par
In GAN models, it is nearly intractable to explicitly trace the evidence of probability distribution of the generated and original data samples. The loss function indirectly estimates the difference between the underlying probability distributions of the generated and target data samples through a divergence measure \cite{f-divergence-GAN}. Hence, loss function significantly impacts the study of GAN research in various ways. As per the existing literature, GAN models are much more promising for the VC task than traditional models.
\par
Likewise in GAN-based VC frameworks speech features are extracted from a source and a target speaker's speech data. The extracted speech features of the source and target speakers are then fed to the generator and the discriminator in terms of speech feature vectors, respectively \cite{GAN-Based-Zero-Shot-VC}. The generator converts the source speaker's speech features $\textbf{x}$ into a particular target speaker's speech features represented by $\textbf{\^y}$. On the other hand, the discriminator is trained with the target speaker's real speech features represented by $\textbf{y}$. The discriminator distinguishes the converted speech features ($\textbf{\^y}$) from real speech features ($\textbf{y}$) with the help of the loss function. In general, CEL is widely used to train a GAN model. The parameters of the generator and the discriminator are updated during the training process through maximization and minimization of the loss function. After achieving the equilibrium state of the GAN model, the vocoder reconstructs the audible speech from the converted speech features without changing the linguistic contents.
\par
\subsection{Types of VC Processes}
The process of VC can be categorized into different types depending on the mapping mechanism employed and the content of the considered speech dataset. Fig. \ref{fig:thesis-001} illustrates the types of VC processes. The preceding literature suggests that the current VC research favours DL-based models or, more specifically, deep generative models due to their effectiveness and superior performance compared to traditional statistical models \cite{Berrak_Sisman, GAN-Survey-2}. Among the variants of deep generative models, GAN models are mostly utilized to perform the automatic VC task for their efficient style transfer capability over other generative models, such as VAEs \cite{A-Survey-VC}. 
\par
\subsubsection{Parallel and Non-parallel VC }
\begin{figure*}[t]
    \centering
    \includegraphics[height=5cm, width=14.5cm]{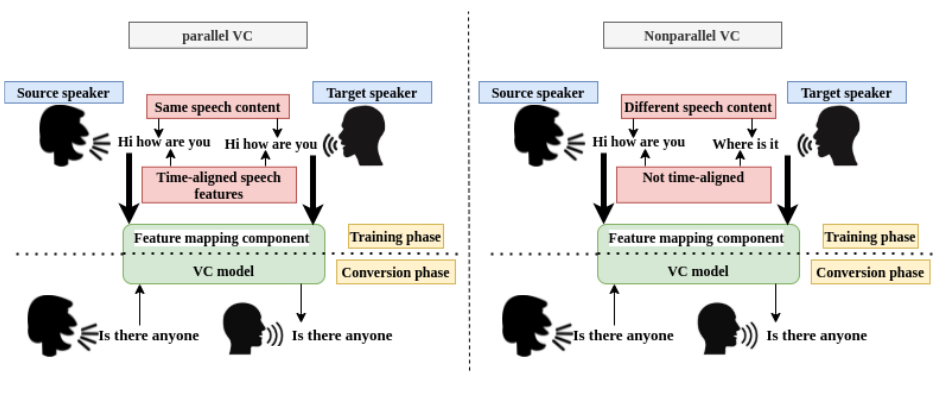}
    \caption{Schematic overview of the parallel VC and non-parallel VC process}
    \label{fig:parallel-and-non-parallel-vc}
\end{figure*}

Based on the linguistic contents of the considered speech dataset, the VC models can be categorized into two main groups: parallel VC and non-parallel VC. In parallel VC, the training dataset comprises similar speech or phonetic content, such as identical spoken words, from different speakers that are time-aligned. DTW \cite{DTW} algorithm is also used in parallel speech data to align utterances. This alignment enables a parallel VC system to readily map the vocal characteristics of source speakers to those of target speakers, facilitated by the precise alignment of linguistic content at the frame level. On the other hand, non-parallel VC systems involve dissimilar linguistic content of source and target speakers. This dissimilarity leads to challenges in effectively mapping source speakers' vocal features to the target speakers' vocal features. As a result, parallel VC systems exhibit superior performance compared to their non-parallel counterparts. However, obtaining parallel speech content in real-world scenarios is challenging, which makes non-parallel VC more realistic despite its inherent complexities \cite{Berrak_Sisman}. The schematic overview of the parallel VC and non-parallel VC processes is shown in Fig. \ref{fig:parallel-and-non-parallel-vc}. 
\par
Voice Conversion Challenge (VCC) 2016 \cite{vcc2016} is a parallel speech dataset consisting of speech samples recorded in US English accents in both male and female voices. On the other hand, VCC 2018 is a non-parallel speech dataset which is also recorded in various English accents (US English) \cite{VCC_2018} (both VCC 2016 and VCC 2018 are mono-lingual speech datasets). As a pioneering deployment of the style transfer GAN within the realm of VC, CycleGAN-VC \cite{CycleGAN-VC} has made significant contributions for both parallel and non-parallel VC. In \cite{CycleGAN-VC}, Kaneko and Kameoka implemented the original CycleGAN framework \cite{CycleGAN2017} for the non-parallel VC. The authors considered MCEP features for speech feature conversion and WORLD vocoder for the reconstruction of audible speech. The model has a residual connection-based 1D gated convolutional neural network (CNN) \cite{CNN} for the generator network, a 2D gated CNN for the discriminator network, and similar loss components as in the original CycleGAN model. However, the model generates buzzy-sounding speech in the case of non-parallel VC due to the over-smoothing problem. In another work, Seshadri et al. \cite{CYCLEGAN_VC4} incorporated PML vocoder \cite{PML} as the first approach in the CycleGAN-VC framework for both speech feature extraction and speech reconstruction. This resulted in a nominal improvement of the generated speech samples. In \cite{CYCLEGAN_VC7}, Du et al. proposed a Spectrum-Prosody-CycleGAN (SP-CycleGAN) framework for cross-lingual VC. The model incorporated a continuous wavelet transform (CWT) \cite{CWT} decomposition technique for $F_0$ modelling instead of a linear transformation of $F_0$. An improved version of CycleGAN-VC was proposed in CycleGAN-VC2 \cite{CYCLEGAN_VC2}. In CycleGAN-VC2, a two-step adversarial loss is used to deal with the over-smoothing problem as observed in CycleGAN-VC. Moreover, an improved generator and a PatchGAN discriminator \cite{Patch-GAN} are incorporated to enhance the effectiveness of the model. As an advancement of the CycleGAN-VC2, in CycleGAN-VC-GP \cite{CycleGAN103}, zero-centred gradient penalties and combined fundamental frequency with the spectrum are considered to ensure the convergence of the GAN model and improvement of the prosody conversion. As a continuation of the development of the  CycleGAN-VC framework, in CycleGAN-VC3 \cite{CYCLEGAN_VC3}, mel-spectrogram conversion was introduced instead of MCEP conversion in the feature mapping stage. This method used time-frequency adaptive normalisation to adjust the scale and bias of the converted features. Likewise, in MelGAN-VC \cite{MelGAN-VC}, spectrogram conversion was incorporated with a siamese network. The model combined a travel loss with adversarial and identity loss to preserve the content information of the speech. In MelGAN-VC, the Griffin-Lim algorithm \cite{Griffin-Lim_Algorithm} was utilized for audible speech reconstruction from spectrograms. MaskCycleGAN-VC \cite{Maskcyclegan-VC} was also developed based on the CycleGAN-VC2 framework, considering mel-spectrogram as the input feature and obtained better results than the preceding models of
the CycleGAN-VC family. The primary difference between MelGAN-VC and MaskCycleGAN-VC is that in MelGAN-VC, the source mel-spectrogram was divided into multiple frames, and the target mel-spectrogram was generated using multiple generators (the same number of generators as the number of frames). In contrast, MaskCycleGAN-VC was developed based on the concept of filling the frame \cite{Filling-black}. Moreover, in  MaskCycleGAN-VC, a MelGAN vocoder \cite{melgan} was utilized for the speech reconstruction process. 
\subsubsection{Mono-lingual and Cross-lingual VC}
\begin{figure*}[t]
    \centering
    \includegraphics[height=5cm, width=14.5cm]{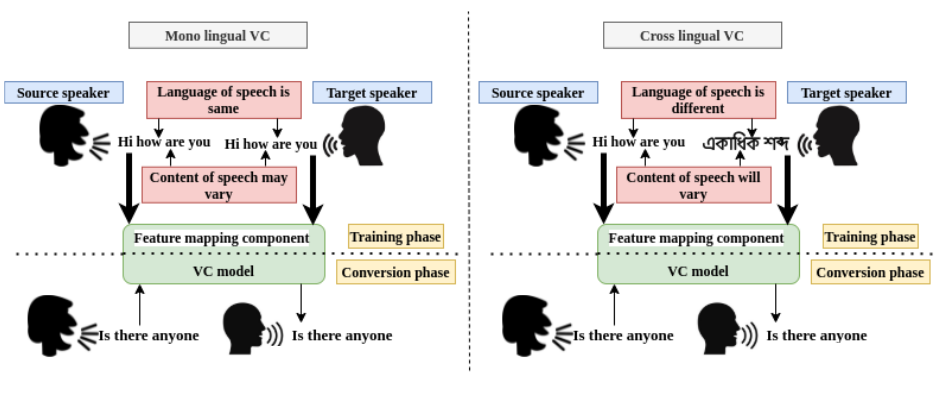}
    \caption{Schematic overview of the mono-lingual and cross-lingual VC}
    \label{fig:monolingual-and-crosslingual-vc}
\end{figure*}

Apart from parallel and non-parallel VC, the classification of VC systems into mono-lingual and cross-lingual VC is determined by the language used in the speech dataset employed for the VC process. In mono-lingual VC, the VC module is trained using a speech dataset that contains the same language for both the source and target speakers. Conversely, in cross-lingual VC, the VC module is trained using speech datasets that consider distinct languages for both the source and target speakers (by default, cross-lingual VC deals with non-parallel speech content) \cite{Cross-Lingual-VC}. Fig. \ref{fig:monolingual-and-crosslingual-vc} shows the schematic overview of the mono-lingual and cross-lingual VC processes. In Fig. \ref{fig:monolingual-and-crosslingual-vc}, the content of speech means phonetic content. 
Since fundamental speech attributes such as syllables, prosody, intonation, and others vary across languages, mapping these speech attributes in cross-lingual VC is more challenging than mono-lingual VC. However, from a practical standpoint regarding the real-time application, cross-lingual VC holds more viability than mono-lingual VC. This is because real-world scenarios often involve situations where speakers communicate in different languages, making cross-lingual VC a more realistic and relevant task despite its existing challenges. VCC 2016, VCC 2018, VCTK \cite{CSTR-VCTK}, and CMU ARCTIC \cite{CMU_Arctic} are mono-lingual datasets (recorded in US English accent). On the other hand, VCC 2020 \cite{VCC_2020} is a widely used cross-lingual dataset recorded in English, Finnish, German, and Mandarin languages. The domain was primarily explored for mono-lingual VC applications during the initial stage of the development of deep-learning-based VC research, due to the unavailability of prominent multi-lingual or cross-lingual speech datasets. However, multiple contributions have been made for cross-lingual VC in the later stage. Cross-lingual VC (CLVC) faces challenges due to misaligned source and target speaker speech contents in the speech dataset. In an attempt to deal with the problems of CLVC, the authors in \cite{PPGS-VC} combined bilingual PPGs (stacking two monolingual PPG vectors) and average modelling to capture language-independent speech features. The basic objective was to associate bilingual PPGs with acoustic features for efficient speech feature mapping and integrate i-vector \cite{i-vectors} for network adaptation. Later on, in another work of deep-learning based CLVC \cite{Cross-Lingual-VC}, the authors proposed a streamlined CLVC framework by utilizing a three-layered neural network for text-independent VC, an autoencoder for source similarity and parallel dataset creation (data augmentation), and a conventional deep neural network (DNN) for feature mapping between source and target. Their proposed model was primarily experimented with a pair of Mandarin-English speakers, and the desired result was obtained compared to the conventional methods. As an initial implementation of GAN for CLVC, Sisman et al. \cite{VAW-GAN} introduced a combined model consisting of variational autoencoding Wasserstein GAN (VAW-GAN) and CycleGAN. The aim of integrating the cycle consistency loss was to enable the simultaneous learning of both forward and inverse mappings, thereby identifying an optimal pseudo pair, even when using cross-lingual training data. The model's performance evaluation experiment was conducted using the CMU ARCTIC and Blizzard Challenge 2010 \cite{Blizzard-Challenge} databases, where their proposed combined model showed better results than individual VAW-GAN and CycleGAN models. In a similar approach \cite{CycleVAE}, Nakatani et al. introduced a novel CLVC method by utilizing a cyclic variational auto-encoder (CycleVAE). It effectively tackled the issue of inaccurate time alignment and language disparities within the source and target speaker's original speech. WaveNet vocoder \cite{WaveNet} improved the conversion quality, showing superior results in English-to-Japanese conversion compared to RNN-based methods. Apart from exploring different variants of the CycleGAN model, the efficacy of the StarGAN framework \cite{StarGAN} was also tested for CLVC by additionally utilizing a VAE model. An $F_0$ injection method was introduced to enhance the $F_0$ modelling during the VC. Both objective and subjective evaluation techniques demonstrated the efficacy of their proposed approach. However, there is a significant research gap regarding the impact of different components of GANs (including loss functions, activation functions and various feature learning mechanisms) on CLVC. This \cite{Shah_Nirmesh2} work explores cross-lingual VC and its challenges with foreign accents affecting speech intelligibility. The authors proposed a novel training scheme with additional linguistic losses that significantly enhance the naturalness and clarity of converted speech between English and Mandarin Chinese. 
\subsubsection{Intra-gender and Inter-gender VC}
\begin{figure*}[t]
    \centering
   \includegraphics[height=5cm, width=14.5cm]{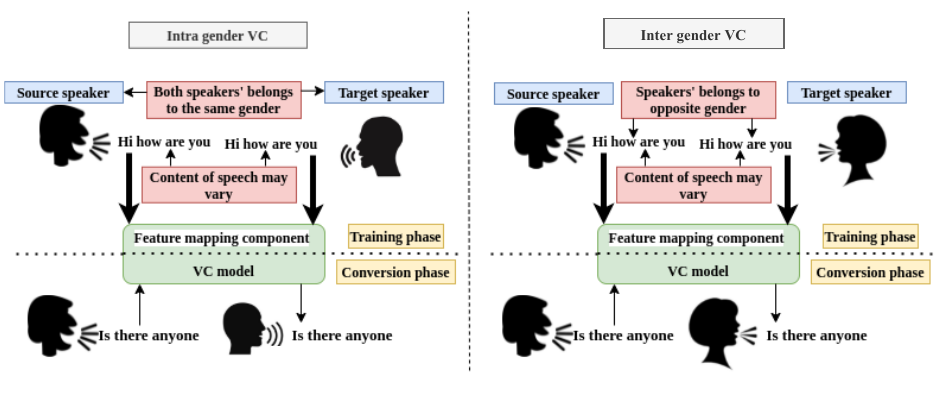}
    \caption{{Schematic overview of the intra-gender and inter-gender VC}}
    \label{fig:Intra-gender-and-inter-gender-VC}
\end{figure*}

VC systems are further categorized into two types: intra-gender and inter-gender or cross-gender VC systems, based on the genders of the speakers participating in the VC process. Speech feature representations such as Mel-scaled power spectrogram, MFCCs, power spectrogram chroma, spectral contrast, and tonal centroid features exhibit variations between male and female speakers. This indicates the significance of the speaker's gender in the VC procedure. As shown in Fig. \ref{fig:Intra-gender-and-inter-gender-VC}, in intra-gender VC, both source and target speakers belong to the same gender, while in inter-gender VC, the source and the target speakers belong to opposite genders. Due to the resemblance in gender-related speech features, mapping speech characteristics is more straightforward in intra-gender VC than in inter-gender VC. Nevertheless, inter-gender VC models demonstrate greater versatility as they can be employed for intra- and inter-gender VC tasks. Benchmark datasets for voice conversion typically include recordings in both male and female voices, making them suitable for both intra and inter-gender VC.  
\par
Most of the developments in VC research consider both intra-gender and inter-gender VC conditions to evaluate the robustness of a model. However, numerous significant models have emerged which are specifically aimed at developing inter-gender VC frameworks to tackle its challenges and several of these models are thoroughly discussed here. Initially, statistical models were employed to address the problems of inter-gender VC. For instance, in \cite{Cross-gender-VC}, the approach involved using the Gaussian mixture conversion (i.e., GMM) model with pre-training on averaged speaker background models, coupled with constant $F_0$ ratio transformation using the WORLD vocoder to enhance cross-gender conversion quality. Likewise, in \cite{Hidden-Markov-Model-VC}, an HMM model was used based on cepstral coefficients conversion for VC by extracting features from similar sentences spoken by source and target speakers (i.e., parallel speech content). The system also utilized pitch ($F_0$) and cepstral coefficients for conversion. Many recent studies on inter-gender VC incorporated hybrid models, e.g., in C-CycleTransGAN \cite{C-CycleTransGAN}, authors introduced the CycleGAN framework alongside a transformer module and a condition embedding network for inter-gender VC. Initially, the model undergoes pre-training via a self-supervised learning mechanism for single-gender voice reconstruction. Subsequently, it is fine-tuned for cross-gender VC, with conditions set to male-to-female or female-to-male VC. In another work on deep learning-based inter-gender VC \cite{inter-gender-vc}, the authors proposed a novel neural architecture for manipulating voice attributes like gender and age inspired by the fader network \cite{Fader-Networks}. It disentangles speech signal information into interpretable voice attributes through adversarial loss minimization, enabling adjustment during VC. Evaluation of inter-gender VC using the VCTK dataset yields promising results, demonstrating the model's ability to learn gender-independent representations. 
\subsubsection{One-to-one and Many-to-many VC}
\begin{figure}[htbp]
    \centering
    \includegraphics[height=5.85cm, width=9.85cm]{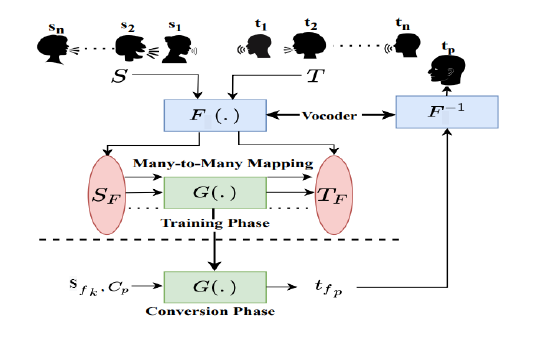}
    \caption{{Basic framework of a typical many-to-many VC system}}
    \label{fig:1-thesis-Intro}
\end{figure}

Based on the type of mapping mechanism employed in the VC process, the VC systems are further divided into two categories: one-to-one and many-to-many VC systems. In one-to-one VC, the VC model can transform the vocal characteristics of a single source speaker into those of a specific target speaker. In contrast, many-to-many VC systems enable the VC model to convert the vocal traits of multiple source speakers into those of multiple target speakers concurrently during the VC training process. Thus, in contrast to one-to-one VC, the inference stage of a many-to-many VC can generate audible speech with the vocal traits of a specific speaker, provided there is a reference speech and a known speaker's identity. However, modeling many-to-many VC systems that can effectively adapt multi-domain speech feature mapping is generally more complex than one-to-one VC, depending on real-time application scenarios. While most VC models have traditionally been designed for one-to-one VC, as discussed in the preceding subsections, recent advancements have primarily emphasized the development of many-to-many VC systems. Mathematically, a typical many-to-many VC system can be explained more precisely. In Fig. \ref{fig:1-thesis-Intro}, 
the basic framework of a typical many-to-many VC system is shown. {Fig. \ref{fig:1-thesis-Intro} can be described using mathematical formulations shown below:
\begin{equation}
\begin{split}
{{X}=F({S})},\\
{{Y}=F({T})},
\label{eq:Equation5}
\end{split}
\end{equation}}where ${{S}}$ and ${{T}}$ indicate all ${n}$ source and target speakers' speech samples, i.e., ${{\{\mathbf{{s}}_i\}}_{i=1}^n}$${\in}$${{{S}}}$ and ${{\{\mathbf{{t}}_j\}}_{j=1}^n}$${\in}$${{{T}}}$. Here ${\mathbf{{s}}_{i}}$ and ${\mathbf{{t}}_{j}}$ basically means the ${i^{th}}$ and ${j^{th}}$ speech sample of source and target speaker respectively. Given ${{S}}$ and ${{T}}$ to the speech feature extraction function ${F(.)}$, the speech features corresponding to all the source and target speaker speech samples are extracted. Therefore, ${{X}}$ and ${{Y}}$ denotes all the extracted speech features belonging to each of the ${n}$ source and target speakers (i.e., ${{\{{{{\mathbf{x}}_{i}}{}}\}}_{i=1}^n}$${\in}$${{{X}}{}}$ and ${{\{{{{\mathbf{y}}_{j}}{}}\}}_{j=1}^n}$${\in}$${{{Y}}{}}$, where ${\mathbf{{x}}_{i}}$ and ${\mathbf{{y}}_{j}}$ means speech features of ${i^{th}}$ and ${j^{th}}$ source and target speaker respectively). Then, the speech feature mapping function
${{Y}{}=G({X}{})}$ maps multiple source speakers' speech features to that of multiple target speakers following the many-to-many mapping framework in the training phase.
\par
In the conversion phase, the trained function ${G(.)}$ is deployed to transform the speech features ${{\mathbf{x}}_{p}}$ of any ${{p}^{th}}$ source speaker to that of any ${{q}^{th}}$ target speaker given the identity of the ${{q}^{th}}$ speaker as ${c_q}$ to the function ${G(.)}$ as shown next:
\begin{equation}
\begin{split}
{{\mathbf{\hat{y}}{}}_{q}=G ({\mathbf{x}{}}_{p},c_q).}
\label{eq:Equation7}
\end{split}
\end{equation} Finally, in the speech reconstruction phase the function ${F^{-1}(.)}$ reconstructs the audible speech ${{\mathbf{\hat{t}}_{q}}}$ from ${{\mathbf{\hat{y}}_{q}}{}}{}$ as shown in Eq. (\ref{eq:Equation8}):
\begin{equation}
\begin{split}
{{\mathbf{\hat{t}}_{q}=F^{-1}({{\mathbf{\hat{y}}_{q}}{}}{}).}}
\label{eq:Equation8}
\end{split}
\end{equation}
\par
Based on the existing literature on VC, it is evident that most of the works in the field of STS synthesis have been developed for one-to-one VC types. However, in \cite{StarGAN-VC}, Kameoka et al. adopted the framework of the StarGAN model \cite{StarGAN} (designed for multi-domain image synthesis tasks) for the many-to-many VC as an initial approach for the development of GAN-based many-to-many VC models, which was termed as StarGAN-VC. The model consists of a generator, a discriminator, and a domain classifier, based on nearly the similar network architecture of CycleGAN-VC \cite{CycleGAN-VC}. The model utilized a WORLD vocoder analyzer for speech feature analysis and audible speech reconstruction. The model provided better results than a VAEGAN-based \cite{VAE-GAN-Framework} VC model in terms of subjective evaluation. However, one major limitation of the StarGAN-VC model was that its performance was evaluated only for VCC 2018 unilingual speech data. Moreover, the performance of the generated samples was only assessed for two subjective cases. Later, in the advancement of the StarGAN-VC model, i.e., in StarGAN-VC2 \cite{StarGAN-VC2}, it was observed that the StarGAN-VC generated samples obtained less evaluation scores in terms of the objective evaluation metrics. Although in StarGAN-VC2, the proposed conditional instance normalizer (CIN) \cite{CIN} improved the objective evaluation scores of the generated samples, StarGAN-VC2 was also evaluated with only VCC 2018 unilingual speech data. This limited the robustness of both StarGAN-VC and StarGAN-VC2 for multi-lingual non-parallel VC. In \cite{PSR-StarGAN}, Li et al. proposed the PSR-StarGAN-VC model, which was incorporated with a perceptual loss function to capture the high-level spectral features during the source-to-target domain feature mapping.
Moreover, residual connection-based architecture and batch normalization are added to the generator network for better feature learning. The samples generated using PSR-StarGAN-VC were better than those from the StarGAN-VC model for the VCC 2018 speech dataset. However, the model was only evaluated in terms of subjective evaluation metrics using the unilingual speech data. In \cite{StarGANv2-VC}, Li et al. proposed StarGANv2-VC, which included perceptual loss and feature-specific loss to the adversarial loss function to increase the speaker similarity of the generated samples. The model was tested on the VCTK, JVS \cite{JVS-dataset}, and emotional speech datasets \cite{ESD} and obtained better results than AUTO-VC \cite{Auto-VC}. However, the model was not compared against other SOTA GAN-based many-to-many VC models. Also, it was not evaluated on a multi-lingual speech dataset. In \cite{F0-Consistent}, Qian et al. proposed $F0$ conditioned AUTOVC to capture the prosodic information during the source to the target domain mapping. Here, the normalized quantized $\log F0$ feature was used to maintain the $F0$ consistency in the generated samples. As a result, the model performed better than StarGAN-VC in terms of subjective evaluation. In \cite{Cycle-Consistent-VAE}, Yook et al. proposed a cycle-consistent variational autoencoder (CycleVAEWGAN) integrated with the Wasserstein loss \cite{Wasserstein-loss} and multiple decoders for many-to-many VC. The model was comprised of an encoder, a decoder, and a discriminator. However, the model was compared against only the variational autoencoding Wasserstein GANs (VAEWGANs) \cite{VAEWGAN} with the VCC 2018 speech dataset. In \cite{Residual-StarGAN-VC}, the authors proposed a residual StarGAN-VC model by incorporating a residual mapping to the framework of StarGAN-VC \cite{StarGAN-VC} to leverage the shared linguistic content between the source and the target features. The residual mapping consisted of identity shortcut connections from the input to the output of the generator network. In \cite{F0-Consistent}, $F0$ consistent many-to-many non-parallel VC was proposed to improve the quality of the generated speech samples in terms of the prosodic information. In \cite{F0-Consistent}, the authors concluded that the normalized quantized $\log F0$ feature transformation technique improved the naturalness of the generated speech samples.
\par
The many-to-many VC mechanism can be modified to function seamlessly in various VC settings, including any-to-one VC  \cite{Any-to-One}, any-to-any VC \cite{Any-to-Any}, and similar configurations.
\par
\subsubsection{Emotional VC}
Apart from the conventional types of VC, another domain of VC, named emotional VC, has recently emerged. Emotional VC refers to altering the emotional characteristics of a person's speech as per some target emotion while retaining the actual content of the input speech. Emotional VC takes the traditional process of VC a step further by focusing on the emotional aspects of speech. The mapping function is trained with parallel speech content in different emotion conditions during emotional VC. In the conversion phase, the emotional condition of the source speaker's speech gets altered to that of a specific target emotional condition. Speech features like pitch, prosody, energy, duration, etc., play a significant role in capturing the state of emotion from a given audible speech data. In essence, emotional VC offers a powerful tool for altering the emotional tenor of spoken communication, enabling applications in fields ranging from entertainment and assistive technology to human-computer interaction. Most of the existing emotional VC datasets, such as the emotional speech dataset (ESD) \cite{ESD}, contain emotions such as neutral, happy, angry, sad, surprised, etc. Emotional VC also encompasses scenarios of non-parallel multilingual VC when the source and target speech have distinct linguistic backgrounds or non-parallel or multilingual content. To carry out the generative model-based emotional VC, in \cite{EVC-1}, Zhou et al. introduced an encoder-decoder-based VAW-GAN for spectrum and prosody mapping, using CWT \cite{CWT} for prosody conversion and explored the impact of F0 feature for improving the emotion conversion. Results demonstrated significant results for both seen and unseen speakers. By extending this work in \cite{EVC-2}, the authors incorporated a pre-trained speech emotion recognition (SER) model to transfer the emotional style during both training and runtime inference. The improvisation made to the baseline model led to superior performance considering unseen speakers. Following a similar approach for the many-to-many VC type, Du et al. \cite{EVC-3} utilized the StarGAN-based framework conditioned on emotional style encoding obtained from a pre-trained SER model. The model worked as a joint emotional style conversion and speaker identity conversion framework referred to as JES-StarGAN and significantly improved overall performance. As an extension of the StarGAN-based VC work, in \cite{EVC-4}, the authors introduced a StarGANv2-VC model for emotional VC. The StarGANv2-VC model was utilized to explore its adaptability in English emotional VC across a variety of speakers and emotions and appeared as an effective framework for voice pitch conversion across emotions. The model showed lower efficiency in multi-emotion to multi-speaker conversion, highlighting the need for further research. For cross-lingual emotional VC, emotion transfer GAN (ET-GAN) \cite{ET-GAN} introduced a cycle-consistent GAN model by utilizing the earth-mover (EM) distance \cite{Earth-Mover-Distance} along with a gradient penalty term. Despite advancements, integrating emotion into speech remains challenging. This \cite{Shah_Nirmesh1} work presents an emotional voice conversion (EVC) system that converts emotions in speech while preserving linguistic content, addressing unseen speaker-emotion combinations. The authors enhanced the StartGANv2-VC architecture with dual encoders and proposed a virtual domain pairing (VDP) training strategy, evaluated using a Hindi emotional database. In \cite{Shah_Nirmesh3}, the authors present a method for emotional speech synthesis that generates speech with a mixture of emotions in real-time by measuring the relative differences between emotional speech samples. They integrated this approach for effective modelling, synthesis, and evaluation of mixed emotions in speech, which was not explored before. Here \cite{Shah_Nirmesh4}, the authors introduced an emotional voice conversion (EVC) method that disentangles speaker style from linguistic content to explicitly control emotion intensity, using an emotion encoder and incorporating classification and similarity losses, with evaluations confirming its effectiveness. 

\subsubsection{Disordered speech-to-normal VC}
Disordered speech-to-normal VC is a technology or process that aims to transform speech characterized by various speech disorders into clear and intelligible normal speech. Speech disorders can include conditions like stuttering, articulation disorders, dysarthria, and other speech impediments that can make communication difficult for individuals. The process of disordered speech-to-normal VC typically involves using advanced speech processing techniques, deep learning algorithms, and speech synthesis technologies. In disordered speech-to-normal VC, the speech data includes disordered speech and corresponding examples of normal speech. The mapping module learns the mapping between disordered and normal speech patterns in the training phase. In the conversion phase, the trained model generates synthesized speech that corrects the speech features extracted from disordered speech samples. EasyCall corpus \cite{easycall} dataset is a well-known dysarthric speech dataset (contains parallel speech in dysarthric and normal speakers' voices). The generated speech samples often pass through a post-processing step, enhancing the naturalness and ensuring that the generated speech samples are more similar to normal human speech. Disordered speech-to-normal VC systems have the potential to significantly improve the quality of life for individuals with speech disorders, as they enable better communication and social interaction. These systems are still an active area of research and development, with ongoing efforts to improve their accuracy and usability. In an attempt to deal with the data scarcity problem of dysarthric speech, the authors in \cite{Dysarthria-VC} introduced a novel method for dysarthria VC that utilized the data augmentation strategy. It combined the Tacotron2 TTS synthesis \cite{NeuralTTS} model and the StarGAN-VC \cite{StarGAN-VC} architecture to generate a substantial corpus resembling both the target and patient-like speech data by reducing the need for extensive recording. The generated samples were evaluated using Google ASR \cite{Google-ASR} metrics and a listening test, and an improvement in the speech intelligibility for dysarthria patients was observed using their proposed system. In \cite{Dysarthria-VC-1}, another approach for the pathological speech data-based VC was proposed, and the authors made prominent contributions to capture the dysarthric speech traits using voice transformer network (VTN) \cite{VTN}, a transformer-based seq2seq model. Due to the data scarcity problem, techniques like TTS pre-training and many-to-one training with normal speakers' data were employed for effective VC. Apart from GAN and TTS, encoder-decoder-based models and auto-encoder models are also used for dysarthric VC. In a related study \cite{Dysarthria-VC-2}, distinct audio-transcript representations via speaker and phoneme recognition encoders alongside an auxiliary reference encoder were employed for the dysarthric speech VC method. This work showed that reconstructing speech from combined linguistic and speaker representations significantly improved conversion quality. In addition to this, the authors of \cite{Dysarthria-VC-3} proposed a novel approach to pathological speech synthesis by customizing the speaker-related features of the existing pathological speech samples to new speakers' characteristics. It demonstrated reasonable naturalness, using a dysarthric speech by utilizing an autoencoder-based technique, particularly for high-speaker intelligibility. However, the preceding works suggest a huge scope for further improvement in the naturalness of the generated dysarthric speech samples for diverse speech contents, specifically in a multi-lingual setting. This \cite{Shah_Nirmesh7} study presents a novel Encoder-Decoder GAN (E-DGAN) for converting multiple types of pathological voices to personalized normal speech, significantly improving intelligibility and content similarity and outperforming five existing methods.  

\subsubsection{Some Recent GAN-based VC Works}
Several recent works have addressed the key challenges in the VC research domain and proposed significant contributions. One line of research introduced a new speech dataset comprising recordings in distinct Indian regional languages alongside their respective English accents. Additionally, efforts have been made to tackle the issue of limited data availability in low-resource languages and pathological speech contexts by employing GAN-based data augmentation techniques. With a focus on generating speech data characterized by high speaker similarity and enhanced speech quality, different benchmark VC datasets were utilized to validate the effectiveness of proposed GAN-based VC models across multiple prominent VC tasks. Key advancements include the development of novel GAN architectures, enhancements in feature learning strategies, and the incorporation of specialized loss functions designed to capture critical speech features, resulting in improved speech quality compared to existing SOTA GAN-based VC models. This discussion exclusively includes our research efforts aimed at addressing the mentioned issues and challenges. The main contributions are summarized as follows:
\par
\begin{itemize}
\item {{ALGAN-VC: An Adaptive Learning based GAN Model for One-to-One VC}}
\par
This work introduced an adaptive learning-based GAN model named ALGAN-VC for one-to-one voice conversion (VC) \cite{ALGAN-VC}. The ALGAN-VC model integrated several substantial contributions aimed at enhancing the quality of the generated speech samples in terms of both objective and subjective evaluation metrics \cite{VC-Evaluation-Metrics}. The architecture of the generator network incorporated a dense residual network to improve the feature learning capability of the model. An adaptive learning approach was proposed for efficient feature learning during loss computation, based on an activation function selection mechanism. A boosted learning rate strategy was also introduced to further enhance the learning capability of the model. Additionally, a new speech dataset encompassing English and various Indian regional language accents was prepared to evaluate the model's performance for multilingual VC. Furthermore, the evaluation was extended to emotional VC tasks to comprehensively examine the model's capabilities across different VC types.
\end{itemize}

\begin{itemize}
\item {{FLSGAN-VC: A Feature Specific
Loss Function based Self-Attentive
GAN Model for VC}}
\par
In this work, a GAN-based VC model named FLSGAN-VC was proposed \cite{FLSGAN-VC}, incorporating a self-attention mechanism \cite{self-attention} within the generator network and a modulation spectra distance loss as a feature-specific loss function that computed the local structural differences between the source and target domains. The self-attention mechanism in the generator network helped to capture the formant distribution of the mel-spectrogram more effectively. Meanwhile, the incorporation of the modulation spectra distance loss aimed to assist the model in generating speech samples with high speaker similarity by reducing the gap between the synthesized and real speech samples.
\end{itemize}

\begin{itemize}
\item {{RNCapsGAN-VC: Region Normalized Capsule Network Based GAN for Non-parallel Voice Conversion}}
\par
To enhance the naturalness of the generated speech samples in non-parallel VC scenarios, an improved GAN-based model, termed as RNCapsGAN-VC \cite{RNCaps-GAN-VC} was developed. The model integrated the region normalization \cite{Region-Normalization} technique in the generator to address covariate shift and a Capsule Network (Caps-Net) \cite{Capsule-Network} based discriminator to better capture formant structures, using a self-attention-based routing mechanism for improved feature representation. Evaluated on multiple non-parallel VC datasets through objective and subjective metrics, the proposed model demonstrated superior performance compared to the SOTA MaskCycleGAN-VC \cite{Maskcyclegan-VC}, achieving higher speaker similarity and speech quality while requiring fewer training iterations. 
\end{itemize}

\begin{itemize}
\item {{FID-RPRGAN-VC: Fréchet Inception Distance Loss-based Region-wise Position Normalized Relativistic GAN for Non-Parallel  VC}}
\par
 In this work, an improved GAN model named FID-RPRGAN-VC \cite{FID-RPRGAN-VC} was proposed for non-parallel voice conversion (VC) to enhance the naturalness of the generated speech samples. The improved GAN model was integrated with a region-wise positional normalization technique in the generator, a relativistic mechanism-based discriminator, and a Fréchet inception distance (FID) \cite{FID} based loss function as the generator loss. The objective of employing the region-wise positional normalization technique was to independently compute normalization statistics at all spatial positions while reducing internal covariate shift. Additionally, an improvised version of the gated linear unit (GLU) \cite{GLU}, called the Gaussian error gated linear unit (GEGLU) \cite{GEGLU}, was incorporated to further improve the model's performance. The inclusion of a relativistic discriminator aimed to trace the similarity between the latent representations of real and generated mel-spectrograms. Moreover, the Fréchet inception distance metric was used as a loss function to efficiently compute the difference between source and target distributions.
\end{itemize}
\begin{itemize}
\item {{CLOT-GAN-VC: Collective Learning Mechanism based Optimal Transport GAN Model for Non-Parallel  VC}}
\par
To learn the speech feature information efficiently using a single-generator multi-discriminator learning strategy, a novel collective learning mechanism-based optimal transport GAN model, termed CLOT-GAN-VC, was developed \cite{ClotGAN-VC}. This model leveraged a unique multi-discriminator learning mechanism comprising the vision transformer \cite{ViT}, conformer \cite{Conformer}, and deep convolutional neural network (DCNN) \cite{DCNN} architectures. The objective of integrating multiple discriminators was to enhance the comprehension of the formant distribution of mel-spectrograms through a collective learning framework. Additionally, an optimal transport loss function was employed to accurately bridge the gap between the source and target data distributions, utilizing the principles of optimal transport theory \cite{OT-Theory}.
\end{itemize}
\begin{itemize}
\item {{GLGAN-VC: A Guided Loss-based GAN Model for Many-to-Many VC}}
\par
To execute the mapping of speech features efficiently within a many-to-many voice conversion (VC) paradigm, a GAN model framework with a guided loss named GLGAN-VC \cite{GLGAN-VC} was designed to improve the quality of the generated speech samples, focusing on architectural enhancements and the integration of alternative loss functions. The proposed approach included a pairwise downsampling and upsampling generator network for effective speech feature mapping in multi-domain VC tasks. Additionally, a feature mapping loss was incorporated to preserve content information, and a residual connection-based discriminator network was introduced to enhance the learning process. A guided loss function was employed to efficiently compute differences between the source and target speakers' deep feature representations in latent space, and an enhanced reconstruction loss was proposed to achieve better preservation of contextual information.
\end{itemize}

\section{VC Datasets}
The VC research community commonly utilizes the parallel dataset VCC 2016 (US English) \cite{vcc2016}, non-parallel dataset VCC 2018 (US English) \cite{VCC_2018}, and multilingual dataset VCC 2020 (English, Finnish, German, and Mandarin languages) \cite{VCC_2020} for VC studies. VCC 2016, VCC 2018, and VCC 2020 are organized with a division into training and testing sets. The details of the VC datasets are provided in Table \ref{table1}.  Table \ref{table1} shows CMU-Arctic \cite{CMU_Arctic} and VCTK \cite{CSTR-VCTK} datasets contain parallel content. However, non-parallel settings are often considered for both datasets by sampling disjoint utterances. CMU-Arctic and VCTK datasets are specifically not divided into train and test sets like VCC datasets; instead, these two datasets can be partitioned into 70:30 or 80:20 ratios for training and testing based on requirements. In addition to the conventional VC datasets, the ESD dataset \cite{ESD} is a multi-lingual dataset recorded in native English and native Mandarin speakers' accents in their respective languages for emotional VC. It has parallel speech contents and considers five emotion classes (neutral, happiness, anger, sadness, and surprise). The ESD dataset can be utilized in a cross-lingual VC scenario as well. As depicted in Table \ref{table1}, the easycall dataset \cite{easycall} comprises dysarthric speech recordings in the Italian language, featuring parallel speech content. It encompasses both recordings from healthy or normal speakers and those with dysarthria. This dataset finds frequent applications in various domains, including speech recognition, speaker identification, and voice conversion.
\begin{table*}[htbp]
\centering
\caption{Summary of the VC datasets}\vspace{0.3cm}
\resizebox{0.8\textwidth}{!}{%
\begin{tabular}{cccccccc}
\hline
\multirow{2}{*}{\textbf{Dataset}} & \multirow{2}{*}{\textbf{Language}} & \multicolumn{2}{c}{\multirow{2}{*}{\textbf{Type}}} & \multicolumn{2}{c}{\textbf{\#Speakers}} & \multicolumn{2}{c}{\textbf{\#Speech Samples}} \\ \cline{5-8} 
 &  & \multicolumn{2}{c}{} & \textbf{Male} & \textbf{Female} & \textbf{Training} & \textbf{Testing} \\ \hline
VCC 2016 & mono lingual & \multicolumn{2}{c}{parallel} & 4 & 4 & 162 & 54 \\ \hline
\multirow{2}{*}{VCC 2018} & mono lingual & \multicolumn{2}{c}{parallel} & 4 & 4 & 81 & 35 \\ \cline{2-8} 
 & mono lingual & \multicolumn{2}{c}{non-parallel} & 4 & 4 & 81 & 35 \\ \hline
\multirow{3}{*}{VCC 2020} & \multirow{2}{*}{mono lingual} & \multicolumn{2}{c}{parallel} & 4 & 4 & 20 & 25 \\ \cline{3-8} 
 &  & \multicolumn{2}{c}{non-parallel} & 4 & 4 & 50 & 25 \\ \cline{2-8} 
 & cross lingual & \multicolumn{2}{c}{non-parallel} & 4 & 6 & 70 & 25 \\ \hline
CMU -Arctic & mono lingual & \multicolumn{2}{c}{parallel} & 2 & 2 & \multicolumn{2}{c}{4528} \\ \hline
VCTK & mono lingual & \multicolumn{2}{c}{parallel} & 47 & 53 & \multicolumn{2}{c}{16262} \\ \hline
ESD & multi lingual & \multicolumn{2}{c}{parallel} & 5 & 5 & 300 & 50 \\ \hline
\multirow{2}{*}{Easycall} & \multirow{2}{*}{mono lingual} & \multirow{2}{*}{parallel} & healthy speech & 14 & 10 & \multicolumn{2}{c}{10077} \\ \cline{4-8} 
 &  &  & dysarthric speech & 20 & 11 & \multicolumn{2}{c}{11309} \\ \hline
\end{tabular}
}
\label{table1}
\end{table*}

\section{Evaluation of Synthetic Speech Obtained from VC Models}
\begin{figure}[htbp]
    \centering
    \includegraphics[height=3cm, width=9cm]{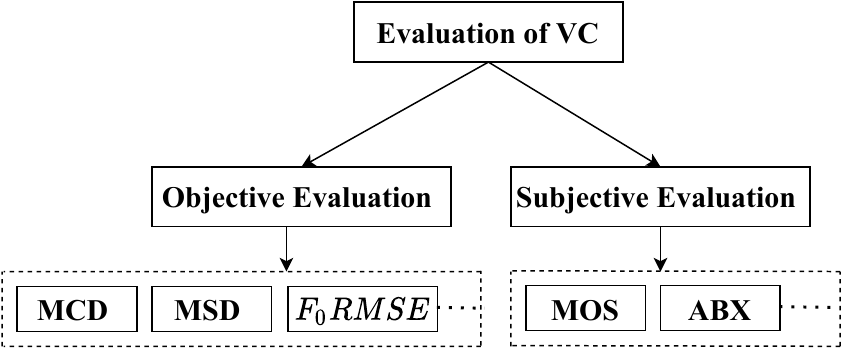}
    \caption{The objective and subjective evaluation metrics for VC model generated speech samples}
    \label{fig:subjective-objective}
\end{figure}

The assessment of speech samples generated by VC models involves the utilization of both objective and subjective evaluation metrics \cite{VC-Evaluation-Metrics}, as shown in Fig. \ref{fig:subjective-objective}. The objective evaluation metrics used for the assessment are mel-cepstral distortion (MCD), modulation spectra distance (MSD), ${\log\hspace{0.001cm}F0\hspace{0.001cm}}$ root mean square error (RMSE), etc. Meanwhile, the mean opinion score (MOS), the ABX test, etc., are utilised for subjective evaluation. Significantly, lower values of MCD, MSD, ${\log\hspace{0.001cm}F0\hspace{0.001cm}}$ RMSE are indicative of high feature similarity, while conversely, a high MOS score indicates high perceptual similarity.
\subsection{Mel Cepstral Distortion (MCD)}
MCD \cite{MCD} measures the global structural differences of the target and converted speech samples. MCD is obtained by considering the squared root differences between the dimension-wise MCC features of the target ${t}$ and the converted speech ${\hat t}$ as shown next:
{\begin{flalign}
&{{{{MCD}}[dB]=\frac{10}{\log\hspace{0.1cm}10} \sqrt{2\sum_{d=1}^{k}{(mcc^{t}_{d}-mcc^{\hat t}_{d})}^{2} }}},
\label{eq:Equation32}
\end{flalign}}where ${mcc^{t}_{d}}$ and  ${mcc^{\hat t}_{d}}$ indicate the ${d}$-th dimensional coefficient of MCC features of ${t}$ and ${\hat t}$, that varies from $1$ to ${k}$ (${k}$ is maximum MCC features).
\subsection{Modulation Spectra Distance (MSD)}
MSD \cite{MSD} is used to measure the local structural differences between the speech samples of target speech ${t}$ and converted speech ${\hat t}$. It is obtained by root mean squared difference between the target and the converted logarithmic modulation spectra ${s{(\mathbf{y})}^{t}}$ and ${s{(\mathbf{y})}^{\hat t}}$ respectively, as depicted below:
{\begin{flalign}
&{{{{MSD}}= \sqrt{\frac{1}{N}\sum_{i=1}^{N}{(s(\mathbf{y})^{t}_{i}-s(\mathbf{y})^{\hat t}_{i})}^{2} }}},
\label{eq:Equation33}
\end{flalign}}where ${i}$ indicates the frame values of logarithmic modulation spectra that vary from $1$ to ${N}$ (${N}$ is the maximum frame value).
\subsection{\texorpdfstring{${\log F0\hspace{0.001cm}RMSE}$}{log F0 RMSE}}
${\log\hspace{0.1cm}F0\hspace{0.1cm}RMSE}$ \cite{F_0RMSE} is used for determining the difference between the prosody or fundamental frequency (i.e. ${F0}$) of target speech ${t}$ and converted speech ${\hat t}$. It is obtained by root mean squared between the target and the converted logarithmic ${F0}$ feature (i.e. ${\log\hspace{0.1cm}F0}$ feature) as shown in Eq. (\ref{eq:Equation34}):
{\begin{flalign}
&{{{{\log\hspace{0.1cm}F0\hspace{0.1cm}RMSE}}= \sqrt{\frac{1}{{N}}\sum_{i=1}^{{N}}{(\log\hspace{0.1cm}{F0}^{t}_{i}-\log\hspace{0.1cm}{F0}^{\hat t}_{i})}^{2} }}},
\label{eq:Equation34}
\end{flalign}}where the frame number $i$ vary from $1$ to ${N}$.
\par
In addition to MCD, MSD, and ${\log\hspace{0.001cm}F0\hspace{0.001cm}}$ RMSE, the signal-to-noise ratio or speech-to-noise ratio (SNR) and peak signal-to-noise ratio (PSNR) are commonly employed to evaluate the quality of generated samples in terms of the presence of background noise. Unlike other objective evaluation metrics, a higher SNR or PSNR value signifies higher quality generated speech samples.
\subsection{Mean Opinion Score (MOS)}
 The MOS \cite{MOS} is collected from human volunteers based on the perceptual quality and speaker similarity of the generated speech samples from the human hearing perspective. In MOS calculation, the values between 1 to 5 are considered, where ``5" is for excellent, ``4" is for good, ``3" is for fair, ``2" is for poor, and ``1" is for bad speech quality. In Eq. (\ref{eq:Equation35}), $R_n$ is the individual rating for a given speech sample, and $N$ is the total number of subjects or human volunteers. 
{\begin{flalign}
&{MOS}=\frac{\sum_{n=1}^{N} R_{n}}{N}
\label{eq:Equation35}
\end{flalign}} 
\subsection{ABX}
The ABX test \cite{abx-test} is a frequently utilized method for the perceptual evaluation of generated speech samples in order to compare and evaluate perceived differences between speech samples. In this test, a listener has three segments: sample A, the first reference, and sample B, the second reference, while X is a randomly picked speech sample from either A or B. The listener's objective is to ascertain whether X matches A or B. Like MOS, this test is crucial in VC research due to the essential role of human perception in assessing the generated speech samples. 

\section{Future Research Directions in VC}
Several promising avenues for future research in this area of VC remain unexplored and hold the potential to significantly advance the field. Below, some key research directions are highlighted that could contribute to a deeper understanding and improvement of the current models and methodologies:
\begin{itemize}
\item Firstly, exploring the correlation between objective and subjective evaluation metrics for accurately assessing the quality of generated speech samples. Investigating how these metrics relate to each other provides valuable insights into refining evaluation methodologies and enhancing the reliability of VC systems.

\item Secondly, integrating visual features of speakers with VC models for audio-visual speech generation holds great potential for creating more immersive and realistic communication experiences. By combining audio and visual information related to the speakers, VC systems can produce synchronized speech with enhanced naturalness and expressiveness.

\item Thirdly, incorporating VC models with TTS synthesis models for real-time speech translation and generation. By seamlessly converting voice characteristics while synthesizing speech from text inputs, these integrated models can facilitate multilingual communication and accessibility.

\item Improving the training mechanisms of VC models to handle low-resource language datasets more efficiently is another crucial area for future exploration. Developing techniques to effectively leverage limited data for training can enable the adaptation of VC technology to a wider range of linguistic contexts.

\item Additionally, focusing on accent conversion for speakers belonging to different dialects within a specific linguistic group can lead to more inclusive and culturally sensitive VC systems. By preserving the linguistic diversity of speakers, accent conversion techniques can promote effective communication across diverse language communities.
\item Efficient VC in the case of mixing multiple language content in a single speech presents a unique challenge that warrants further investigation. Developing models capable of accurately converting voices in multilingual contexts can facilitate seamless communication in diverse linguistic environments.
\item Furthermore, addressing the need for adequate VC for longer ranges of speech data is essential for practical applications such as audiobooks and podcasts. Research in this area can focus on extending the capabilities of VC models to handle extended speech segments while maintaining high quality and coherence.
\item Incorporating speech enhancement models with VC models to improve the results of generated speech data offers opportunities for enhancing the overall quality of VC systems.
\item Moreover, utilizing VC for specific speakers by simultaneously identifying speaker identity from a mixture of speech uttered by multiple speakers can enable personalized communication experiences. Developing speaker identification techniques tailored to mixed speech scenarios can enable targeted voice conversion for individual speakers within a group conversation.
\item Future research in VC could explore the use of diffusion models to enhance speech quality and diversity, along with hybrid approaches that integrate the strengths of both GANs and diffusion processes for more accurate and expressive conversions.
\item Lastly, reducing the latency of VC is crucial for real-time applications such as voice assistants and telecommunication systems. Investigating methods to streamline the processing pipeline and optimize computational efficiency can help minimize latency and improve the responsiveness of VC systems in interactive settings.
\end{itemize}
\section{Conclusion}
This survey has explored the advancements in GAN-based VC technology, highlighting the key challenges and solutions that have shaped the development of more effective VC systems. The survey paper discussed various GAN-based architectures, loss functions, and learning mechanisms that address the primary obstacles in generating realistic speech, preserving linguistic content, and enhancing synthesis quality. The survey reviewed several SOTA GAN-based VC models, each offering unique innovations in architecture and loss functions. These models, evaluated across multiple datasets, demonstrated significant improvements in capturing formant details, contextual nuances, and maintaining speech naturalness in both mono- and cross-lingual VC tasks. Despite these advancements, challenges remain, particularly in aligning speech duration with natural timing in generated samples. Future research could address these issues by incorporating more robust duration modeling techniques.

\bibliographystyle{IEEEtran}
\bibliography{reference.bib}
\end{document}